\documentclass[letterpaper,12pt]{article}
\usepackage{amsmath, amsthm, amssymb, fancyhdr, fancybox, graphicx, url, color, wasysym}
\usepackage{natbib}
\usepackage{sectsty}
\sectionfont{\large}
\usepackage[margin=0.75in]{geometry}
\usepackage{times}

\usepackage{graphicx}

\begin{document}

\title{Calibrating wood products for load duration and rate: A statistical look at three damage models}

\author{Samuel W.K. Wong\\
	Department of Statistics and Actuarial Science, University of Waterloo %, Waterloo ON, Canada
}

\date{February 8, 2020}

\maketitle{}

\begin{abstract}	
Lumber and wood-based products are versatile construction materials that are susceptible to weakening as a result of applied stresses.  To assess the effects of load duration and rate, experiments have been carried out by applying preset load profiles to sample specimens.  This paper studies these effects via a damage modeling approach, by considering three models in the literature: the Gerhards and Foschi accumulated damage models, and a degradation model based on the gamma process.   We present a statistical framework for fitting these models to failure time data generated by a combination of ramp and constant load settings, and show how estimation uncertainty can be quantified.  The models and methods are illustrated and compared via a novel analysis of a Hemlock lumber dataset.  Practical usage of the fitted damage models is demonstrated with an application to long-term reliability prediction under stochastic future loadings. 
\end{abstract}
%To ensure their long-term reliability when placed into service, 
 %This paper studies the effect of load duration and rate on wood products via a damage modeling approach.
%The duration of load and the rate at which it is applied
% A key application of the fitted damage models is to analyze reliability of 
% When fitted, such models
% As an application of the fitted damage models, their us for predicting reliability under stochastic future stresses, so 

% The usage of the fitted damage models is demonstrated with an application to predicting reliability under stochastic future stresses. 
%so 
% models provide a
%
% Damage models permit the reliability of wood products to be analyzed under stochastic load profiles

\section{Introduction}

For nearly 70 years it has been recognized in the literature that the strength of wood changes over time due to applied stresses, as a function of both the duration of the load \citep{wood1951relation} and the rate at which the load is applied \citep{liska1950effect}.  These are known as the ``duration-of-load'' (DOL) and ``rate-of-loading'' (ROL) effects in lumber.  In particular, the gradual weakening of lumber over time must be considered in the construction of wood-based structures, and this factor is governed by appropriate standards to ensure the long-term reliability of structures \citep{astmD681509}.

Experiments have been carried out on lumber specimens to collect empirical data for assessing the extent of the DOL and ROL effects \citep[e.g.,][]{wood1951relation,madsen1982duration,foschi1982load,gerhards1987cumulative,karacabeyli1993rate}.  Briefly, the common idea is to design a load profile $\tau(t)$ under which a sample is to be stressed over time $t \ge 0$, and record the time to failure for each sample specimen.  Under a \emph{ramp load} test we have $\tau(t) = kt$, that is, the load is increased linearly at rate $k$ until the piece breaks.  For a \emph{constant load} test with level $\tau_c$, the load is first increased linearly  $\tau(t) = kt$ for  $0 \le t \le \tau_c / k$, after which the load is held at the constant level $\tau_c$ until either the specimen fails or a preset time $T_1$ is reached, the latter usually being several months or years.
% time $T_1$, which is usually several months or years in duration.
The settings for $k$ in the ramp load test and $\tau_c$ along with $T_1$ in the constant load test are usually varied across different samples.  For the constant load test, when specimens survive the entire duration $T_1$, these survivors can then be ramp loaded to failure for further study, as seen in \cite{gerhards1987cumulative} and \cite{erol1994}.  With the introduction of new engineered wood products such as oriented strand board and cross-laminated timber, DOL is a continued avenue of research as these wood-based products are also expected to be susceptible to DOL effects \citep[e.g.,][]{wang2012durationB,li2016reliability,gilbert2019reliability}.

%ASTM D4761-88  1 minute test, defined as the maximum stress sustained by a piece in a short-term
In the earliest known DOL study, \cite{wood1951relation} used his empirical data to fit a curve relating the time-to-failure and applied load as a percentage of the short-term strength of the piece, and this relationship has come to be known as the Madison curve.  The standard test for measuring short-term strength is a ramp load test with an average duration of 1 minute \citep{D4761:2005lr}, and we denote this short-term strength by $\tau_s$ and the corresponding ramp load rate by $k_s$.  In a similar vein, \cite{karacabeyli1993rate} fitted log-linear relationships in a comprehensive ROL study to predict strength under different ramp loading rates $k$. However, as a practical limitation, such fitted curves do not have the capacity to predict time-to-failure and assess reliability under the arbitrary load profiles $\tau(t)$ that may be encountered when lumber is placed in service.  For these scenarios, models were needed, which we now briefly overview.

%Several models have been proposed and adopted 
A general class of models was developed for this purpose, which we refer to as accumulated damage models (ADMs), see, e.g., \citet{ellingwood1991duration}.  First define $\alpha(t)$ as the damage sustained by a specimen over time, with $\alpha(0)=0$ indicating no damage and $\alpha(T_f) = 1$ when the piece fails at time $T_f$ as a result of applied stress.  Then, the ingenuity of this approach is to model the rate of change in damage $\dfrac{d}{dt} \alpha(t)$ using a differential equation that involves the applied load $\tau(t)$ and various parameters. One early example was proposed by \cite{gerhards1979time} and known as the `US model', which specifies
\begin{equation} \label{US_model}
\frac{d}{dt}\alpha(t)  =\exp\left(-A+B\frac{\tau(t)}{\tau_{s}}\right)
\end{equation}
where $A$ and $B$ can be treated as model parameters or random effects, and $\tau_s$ represents short-term strength --  a key idea being that $\tau_s$ represents the intrinsic strength of a specimen and hence can serve as a kind of covariate or predictor of its long-term strength properties.  With its simplicity and ease of computation, the US model is still widely-used today \citep[e.g.,][]{gilbert2019reliability,wang2019parameter}. 
Care should be taken to ensure that such ADMs are dimensionally consistent and do not depend on the units of measurement used \citep{wong2018dimensional}; for example, in the equation (\ref{US_model}) the left-hand-side strictly should be multiplied by a constant with units `time' so that both sides are unitless.  A more complex ADM was proposed by \cite{foschi1982load} and refined in \cite{foschi1986duration2}, and has become known as the `Canadian model'.  While it has been shown to fit empirical DOL data better than the US counterpart \citep{foschi1986duration2}, it is not straightforward to fit the Canadian model with random effects using standard statistical methods; for example, in \citet{hoffmeyer2007duration} and \citet{kohler2011probabilistic}, simplified versions of the Canadian model with fixed effects were considered instead.
% (We note that in \citet{hoffmeyer2007duration} and \citet{kohler2011probabilistic}, simplified versions of the Canadian model with fixed effects parameters were considered.)
A principled approach to handle the random effects and characterize uncertainty was only recently proposed under a Bayesian framework \citep{yang2019bayesian}.  Finally, in recent work an entirely distinct approach from ADMs was proposed, that uses a gamma process to model degradation with the benefit of having parameters that are easier to interpret \citep{wong2019duration}.

For any chosen damage model, the typical workflow is to fit its parameters based on experimental data.  The fitted model can then be used to predict reliability under a variety of loadings.  However, to the best of our knowledge, only the US model has been previously fitted to data under the three aforementioned test profiles simultaneously, for model calibration as described in \cite{gerhards1987cumulative}.  These test profiles are also illustrated in Figure 1 of \cite{wang2019parameter} and analyzed in the context of the US model.  We define them here, for reference in the sequel as follows:
\begin{enumerate}
	\item[R.] Ramp load test $\tau(t) = kt$, with varying loading rates $k$.  When $k=k_s$, this test determines the short-term strength $\tau_s$.
	\item[C.] Constant load test 
	$$\tau(t) = \begin{cases} kt  \mbox{~~~~for~~~}  0 \le t \le T_0 \\ \tau_c \mbox{~~~~for~~~}   T_0 < t \le T_1 \end{cases}$$
	with varying constant load levels $\tau_c$ and durations $T_1$.  Note that $T_0 = \tau_c / k$ denotes the time required for the load to initially reach $\tau_c$.% in the initial ramp loading.
	\item[RCR.] Ramp load test of the constant load survivors, $\tau(t) = k(t-T_1)$ for $t > T_1$, on specimens that survive to the end of a constant load test.  We may call this the `ramp-constant-ramp load' (RCR) test.
\end{enumerate}
As discussed in \cite{gerhards1987cumulative}, the US model is limited in its ability to adequately fit all the data from the different test scenarios.  The same three test scenarios were used in the Western hemlock experiments described in \cite{foschi1982load} and \cite{erol1994}; however, the RCR data produced from that experiment have not been previously analyzed in a damage modeling framework.

These considerations motivate the main contributions of this paper.  First, we adapt the Canadian model computational framework of \cite{yang2019bayesian} and the gamma process approach of \cite{wong2019duration} to include the fitting of RCR data and R data with varying ROLs.  Second, we then analyze the \cite{foschi1982load} and \cite{erol1994} ramp load, constant load, and RCR data using all three of the US, Canadian, and gamma process models.  Third, we perform a reliability analysis that compares the three fitted models and accounts for estimation certainty.  In doing so, we show how advances in statistical computation can be leveraged in future DOL research, by broadening the suite of available models for corroborating reliability assessments in engineering applications.

The remainder of the paper is laid out as follows.  In section 2, we describe the experiment and data of \cite{foschi1982load}. In section 3 we review the three damage models and statistical approaches for estimation, then adapt the Canadian and gamma process models for handling RCR data.     The model fitting results are shown in section 4.  An illustrative reliability analysis comparing the models is presented in section 5.  We conclude with a brief discussion in section 6.

\section{Experimental data}

The data we analyze in this paper were generated by the experiments first described in \cite{foschi1982load}. Samples of (nominal) 2-by-6 Western hemlock lumber, No.~2 grade or better, were tested under a three-point bending setup with a 3.51m (138 inch) span at a Forintek Canada (the predecessor of today's FPInnovations) laboratory in Richmond, British Columbia, Canada.  The samples were divided into groups for different ramp load and constant-load test profiles, such that the distributions of the modulus of elasticity (MOE) were as similar as possible across the groups.  All strengths and load levels will subsequently be given in units of MPa, along with `pounds per square inch' (psi) as used in the original study where helpful.  The reference loading rate used for determining short-term strength was set as $k_s = 2678$MPa/hour, which corresponds to an average test duration of approximately one minute.  All time units are `hours' unless otherwise specified.

For the ramp load groups, five different load rates $k$ were applied, and these samples are summarized in Table \ref{tab:ramp};  $k$ is expressed in terms of $k_s$,  thus group 3 measures $\tau_s$ since it applies $k=k_s$.  These data were subsequently included in the ROL study by \cite{karacabeyli1993rate}.
%, where the $k$'s are expressed as 
\begin{table}[!htbp]
	\caption{Summary of ramp load data from the Forintek Canada experiment.  Five different rates of loading were used, as indicated in the Rate column in terms of $k_s$.}
	\begin{center}
		\begin{tabular}{c|c|c|c|c}
			Group &   Rate $k$ ($\times k_s$) & sample size &    mean time-to-failure &    mean strength (MPa) \\
			\hline
     		1 & $1.667 \times 10^{-3}$ & 140 & 619 minutes & 44.80 \\
     		2 & 0.0333 & 139 & 31.8 minutes & 46.35 \\
     		3 & 1.0 & 139 & 65.7 seconds & 47.83 \\
     		4 & 30.0 & 139 & 2.13 seconds & 46.59 \\
     		5 & 1500 & 140 & 0.0453 seconds & 48.95
		\end{tabular}
	\end{center}
	\label{tab:ramp}
\end{table}

For the constant load groups, two different constant load levels ($\tau_c$) were applied:  20.68 (3000psi) and 31.02 (4500psi), designed to respectively represent the 5th and 20th percentiles of $\tau_s$ for this size and species.  The rate $k=k_s$ was used in the initial phase for increasing the load to the specified constant load level.  Durations of the constant-load test ($T_1$) ranged from 3 months to 4 years.  At the end of the constant load period, the surviving samples were unloaded and ramp-load tested using rate $k=k_s$ to yield RCR data.  These groups are summarized in Table \ref{tab:const}, where the last three columns indicate the number of failures grouped by time of occurrence: initial ramp load to $\tau_c$, constant load period, and ramp load test of survivors.   At the time of writing of \cite{foschi1982load}, the constant-load test was still in progress in its first year.  The data from the completed constant-load tests were subsequently analyzed in \cite{foschi1986duration2} using ADMs.  The RCR data were finally examined by \cite{erol1994}, where an empirical assessment of strength degradation due to constant-load  was made, but  damage models were not fitted.

\begin{table}[!htbp]
	\caption{Summary of constant load and RCR data from the Forintek Canada experiment. Five different combinations of constant load level and duration were used, as indicated in the $\tau_c$ and $T_1$ columns respectively.}
	\begin{center}
		\begin{tabular}{c|c|c|c|c|c| c }
			& & & & \multicolumn{3}{c}{\# failures during time interval}\\
			Group &   $\tau_c$ & $T_1$ & sample size &  $t \le \tau_c/k_s$  &  $ \tau_c/k_s < t \le T_1 $  & $t > T_1$ \\
			\hline
			6 & 20.68 & 3 mo. & 300 & 17 & 18 & 265 \\
			7 & 20.68 & 4 yr. & 198 & 4 & 42 & 152 \\
			8 & 31.02 & 3 mo. & 98 & 19 & 26 & 53\\
			9 & 31.02 & 1 yr. & 300 & 57 & 97 & 146 \\
			10 & 31.02 & 4 yr. & 101 & 23 & 41 & 37
		\end{tabular}
	\end{center}
	\label{tab:const}
\end{table}
% double check these numbers.

\section{Damage models and their estimation}

In this section we review the US \citep{gerhards1987cumulative}, Canadian \citep{foschi1986duration2}, and gamma process \citep{wong2019duration} models.  For the latter two, we also show how to adapt existing statistical estimation procedures to incorporate failure time data from all three test profiles (R, C, and RCR) used in the experiment. % described in the Introduction.

\subsection{US model}
The US model was originally introduced by \cite{gerhards1979time}, and specifies the damage $\alpha(t)$ for a specimen according to the differential equation (\ref{US_model}).  It is the simplest model of the three, and the only one for which an estimation procedure has been proposed to fit data from the three test profiles \citep{gerhards1987cumulative}.  We summarize this procedure briefly.

First, $\tau_s$ for a piece is assumed to have a log-normal distribution:  $\tau_s = \tau_M  \exp(wZ)$, where $\tau_M$ is set to be the median short-term strength of the sample, $Z$ is a standard Normal random variable, and $w$ is a scale parameter to be estimated.  Then, under each test profile, the differential equation (\ref{US_model}) may be solved analytically using the conditions $\alpha(0) = 0$ and $\alpha(T_f) = 1$ at the time of failure $T_f$ (solutions are listed in Appendix \ref{app:US}).  Given the  failure times $T_f$ in the sample, an iterative weighted nonlinear least squares (NLS) procedure is then used to estimate the parameters $A$, $B$, and $w$ (see Appendix \ref{app:US} for details).

\subsection{Canadian model} \label{est:canadian}

Like the US counterpart, the Canadian model of  \cite{foschi1986duration2} uses an ODE to describe the progression of damage $\alpha(t)$ in a specimen.  We use the parametrization in \cite{yang2019bayesian} which ensures the model is coherent under dimensional analysis: % based on a series expansion

\begin{eqnarray} \label{Can_model}
\frac{d}{dt}\alpha(t) \mu &=& [(a \tau_s )(\tau(t)/\tau_s - \sigma_0)_+]^b
+ [(c \tau_s )(\tau(t)/\tau_s - \sigma_0)_+]^n \alpha(t)
\end{eqnarray}
where $a$, $b$, $c$, $n$, $\sigma_0$ are random effects specific to the piece and assumed to follow lognormal distributions, see Appendix B for details.  Further,  $(x)_+ = max(x, 0)$ and $\mu$ is any constant with units `time', which we set as the time unit used to measure the failure times, namely $\mu = 1$ hour.  Finally, as before $\tau_s$ is the short-term strength, which can be shown to be a function of the five random effects. This model has the feature of a stress ratio threshold $\sigma_0$, in that damage to a piece only occurs when  $\tau(t)/\tau_s  > \sigma_0$, that is, when the applied load exceeds $\sigma_0$ times its short-term strength. It has a total of 10 parameters to be estimated, namely the mean and variance parameters associated with each of the five random effect distributions.

Fitting the Canadian model poses a challenge:  unlike the US model, the time-to-failure does not have analytic solutions under any of the test profiles R, C, or RCR; further, the multiple random effects nested within the ODE precludes the construction of a standard likelihood or objective function for parameter estimation.   In \cite{yang2019bayesian}, a new estimation approach based on approximate Bayesian computation (ABC) was proposed, which fits the model to ramp and constant load data and characterizes uncertainty in the parameters.  Therein, expressions for time-to-failure under R and C profiles in the case $k=k_s$ are provided as implicit solutions to equations involving the random effects.
% A simpler variant of the model, which treats the parameters as fixed quantities, was considered in....

Here, we extend that approach by first deriving expressions for failure time under RCR, as well as for R with different rates of loading  $k$.  Under RCR, the failure time $T_f$ when ramp loading is done at the rate of $k=k_s$ can be found as the solution to the equation
\begin{equation}\label{eq:TfRCRsoln}
H(T_{f}-T_1) - \alpha(T_1)=\frac{\left(a \tau_s \right)^{b}}{\left(c \tau_s \right)^{n(b+1)/(n+1)}}\left(\frac{\mu  (n+1)}{ \tau_s / k_s}\right)^{\frac{b-n}{n+1}}\int_{0}^{-\log H(T_{f} - T_1)}e^{-u}u^{(b+1)/(n+1)-1}du
\end{equation}
where $H(t)$ is the integrating factor
\begin{equation} \label{eq:intfac}
H(t)=\exp\left\{ -\frac{1}{\mu}\left(c \tau_s \right)^{n}\frac{\tau_s}{k_s(n+1)}\left(\frac{k_s t}{\tau_{s}}-\sigma_{0}\right)^{n+1}\right\} .
\end{equation}
Derivation details, along with the expression for $\alpha(T_1)$, the damage accumulated by the end of the constant load period, are presented in Appendix \ref{app:Can}.  Using similar techniques, we also obtain equations for failure time under R with varying ROLs $k$ (see Appendix \ref{app:Can}).
%These also can be expressed as solutions to equations involving the random effects.  The expressions however are rather cumbersome, and so we defer their presentation and mathematical details to Appendix \ref{app:Can}.

Let $\theta$ denote the vector of 10 random effects parameters of interest.  Then the ABC framework from \cite{yang2019bayesian} uses a Markov chain Monte Carlo (MCMC) algorithm to fit the model and can be summarized as follows (the interested reader may refer to that paper for technical details).  First, draw a new value $\theta'$ from a proposal distribution.  Second, use $\theta'$ to simulate a set of failure time data, with the same load profiles and sample sizes as in the real data.  It is within this step that the derived failure time solutions are repeatedly used.  Third, \textit{summary statistics} are calculated from the simulated and real datasets; intuitively, $\theta'$ is a good fit to the data if the differences in summary statistics between the simulated and real datasets are `small'. Fourth, the value $\theta'$ is accepted according to a Metropolis-Hastings rule when the differences are `small enough', and rejected otherwise.  After these MCMC steps are repeated a large number of times, the collection of $\theta'$ values represent the posterior probability distribution of $\theta$ from which parameter estimates and credible intervals (CIs) can be calculated.
%to be estimated from failure time data
% to measure their `similarity'

This ABC framework was shown to provide good fits to R and C datasets  in \cite{yang2019bayesian}:  expressed on the log scale, the summary statistics for R were chosen to be 19 equally spaced quantiles of failure times from $5\%$ to $95\%$; while for C, the summary statistics were 19 equally spaced quantiles from $5\%$ to $95\%$ of the observed failure times, along with the proportion of pieces surviving to the end of the constant-load test.  We extend the ABC framework by defining a corresponding set of summary statistics for RCR data:  the proportion of pieces surviving to the end of the constant-load test, along with 19 equally spaced quantiles from $5\%$ to $95\%$ of $(T_f - T_1)$ for the RCR portion of the test.
%Markov chain Monte Carlo (MCMC) procedure
%At each iteration, a proposal for $\theta$
%Generate data from 
%Using a set of summary statistics to measure how `similar' the simulated dataset is to the real data.
%For ramp, the equally spaced quantiles.
%For constant,
%For RCR
With these adaptations, we are equipped to fit the Canadian model to all of our R, C, and RCR datasets.

\subsection{Gamma process model} \label{est:gamma}

The gamma process model of \cite{wong2019duration} takes an alternative approach to modeling damage:  the accumulation of damage within a specimen is viewed as a stochastic (random) process, rather than a  process determined by an ODE.  Denote this stochastic process representing damage by $Y(t)$ which is non-decreasing over time $t \ge 0$, again scaled such that $Y(0) = 0$ is the initial state with no damage, and $Y(T_f)=1$ at the failure time $T_f$.  Then, as is common in degradation modeling applications, $Y(t)$ is assumed to follow a gamma process:  the damage that accumulates between times $t_1 < t_2$, namely $Y(t_2) - Y(t_1)$, has a gamma distribution with scale parameter $\xi$ and shape parameter $\eta({t_2})- \eta(t_1)$, where $\eta(t)$ is a non-decreasing function over time.  We may call $\eta(t)$ the time-varying shape parameter for the gamma process, which depends on the load history $\tau(t); t\ge 0$.  In this way, the effect of the load profile is captured by $\eta(t)$, while the variability among pieces is captured by $Y(t)$.

First taking the simple case of a constant load $\tau$ applied from time $0$ to $t$, the authors considered modeling the shape parameter via
\begin{eqnarray}\label{eq:eta_tsimp}
\eta(t) = u\, g(t) \times (\tau - \tau^*)_+,
\end{eqnarray}
where $u$ is a scaling parameter, $\tau^*$ is the stress threshold below which no damage occurs, and $g(\cdot)$ is an increasing function due to the DOL effect.
This idea is generalized to handle arbitrary load profiles $\tau(t)$ that vary over time, by defining
%, the authors proposed the following form of $\eta(t)$ to model the DOL effect as a function of $\tau(t)$,
\begin{eqnarray}\label{eq:eta_t}
\eta(t) = u \sum_{i=1}^m g(\tilde{t_i}) \times  \left[(\tau_i -\tau^*)_+ - (\tau_{i-1}-\tau^*)_+ \right],
\end{eqnarray}
where now %$u$ is a scaling parameter, $\tau^*$ is the stress threshold below which no damage occurs, 
$0 = \tau_0 < \tau_1 < \tau_2 < \cdots < \tau_m$ is a sequence of load levels such that $\max \{\tau(t); t\ge0 \} \le \tau_m$, %$g(\cdot)$ is an increasing function, 
and $\tilde{t_i}$ is the total time duration for which $\tau(t) \ge \tau_i$.  Thus for load levels above the stress threshold $\tau^*$, $u (\tau_i - \tau_{i-1})$ captures the contribution of the load increment from $\tau_{i-1}$ to $\tau_i$, and $g(\tilde{t_i})$ captures the DOL effect of that load increment.  It can be seen that (\ref{eq:eta_t}) naturally simplifies to (\ref{eq:eta_tsimp}) in the special case of a constant load $\tau$ from time $0$ to $t$, by setting $\tau_m = \tau$.
%of increasing the load

Standard likelihood-based statistical methods can be applied to this gamma process model, and demonstrated on R and C data in \cite{wong2019duration}.  It is straightforward to extend its applicability to the RCR profile, for example, when $\tau_c = 31.02$ and $k=k_s$ then for the load thresholds $\tau_i = 30$ and $\tau_{i'} = 40$, 
$$\tilde{t_i} = \begin{cases} 0   & \text{for }  t \le 30/k_s \\
 t - 30/k_s  &  \text{for }   30/k_s < t \le T_1 \\
 T_1 - 30/k_s  &  \text{for } T_1 < t \le T_1 + 30/k_s \\
 t - 2 \times  30/k_s &  \text{for }  T_1 + 30/k_s  < t
 \end{cases}$$
and 
$$\tilde{t_{i'}} = \begin{cases} 0   & \text{for }  t \le T_1 + 40/k_s \\
t - (T_1 + 40/k_s)  &  \text{for }    t > T_1 + 40/k_s
\end{cases}$$
are the total time durations for which the loads exceed 30 and 40 respectively, as functions of $t$.

The specific choice of $g(\cdot)$ that governs the effect of load duration is designed to be adaptable.  The failure times in our datasets range from a fraction of a second to several years.  To increase flexibility for fitting this large temporal range of data, we adopt a piece-wise (or broken) power law \citep[e.g.,][]{agnew1992time}:
$g(t) \propto  (t/t_i)^{a_i}$ for $ t_{i-1} < t \le t_i$, where $t_0 = 0$ and $t_1, t_2, \ldots$ is a sequence of time breakpoints with corresponding powers $a_1, a_2, \ldots$ to be estimated.  The constant of proportionality is set to ensure a continuous curve.  We use a standard statistical model selection criterion, namely the BIC \citep{schwarz1978estimating}, to determine the appropriate number of time breakpoints to include in the final fitted model.  Note that the BIC is similar in principle to the well-known AIC \citep{akaike1974new}, but penalizes over-fitting with too many parameters more strongly than AIC.

% AND STATE FORM.  Allow the data fitting process to determine the appropriate number of breaks and BIC.

% results section:  3 models

\section{Model fitting results}
In this section we show the results of applying the model fitting procedures to the experimental data, and then make comparisons between the three models considered.
% the three models considered

\subsection{Assessing goodness-of-fit} \label{sec:GOF}
An important aspect of model fitting involves assessing the goodness-of-fit to the data.  For this purpose, we follow the same procedure for each fitted model.  Using the parameter estimates, we simulate 100,000 failure times from the model for the load profile corresponding to each group listed in Tables \ref{tab:ramp} and \ref{tab:const}.  Then, to facilitate visual assessment across different time scales, we represent each failure time $T_f$ using the corresponding load level sustained at failure, $\tau(T_f)$.  We use these values from the simulated and real data to construct a quantile-quantile (QQ) plot for each group; when the model fits well, the points will largely follow the 45 degree ($y=x$) line.
Further, for quantitative model assessment, we also calculate the value of the likelihood function at the parameter estimates.  For the US and Canadian models the likelihood cannot be explicitly calculated, but a large set of simulated failure times may be used to numerically approximate its value \citep{yang2019bayesian}; to do so, we use kernel density estimation (KDE) on the 100,000 simulated failure times to approximate the probability density of the failure time distribution for each group.  The specific KDE implementation we used is from \cite{botev2010kernel}.
%the simulated failure times are also used to approximate the value of the likelihood function at the parameter estimates.
%To do so, 

\subsection{US model}
\begin{figure}[!htbp]
	\centering
	\includegraphics[scale=0.427]{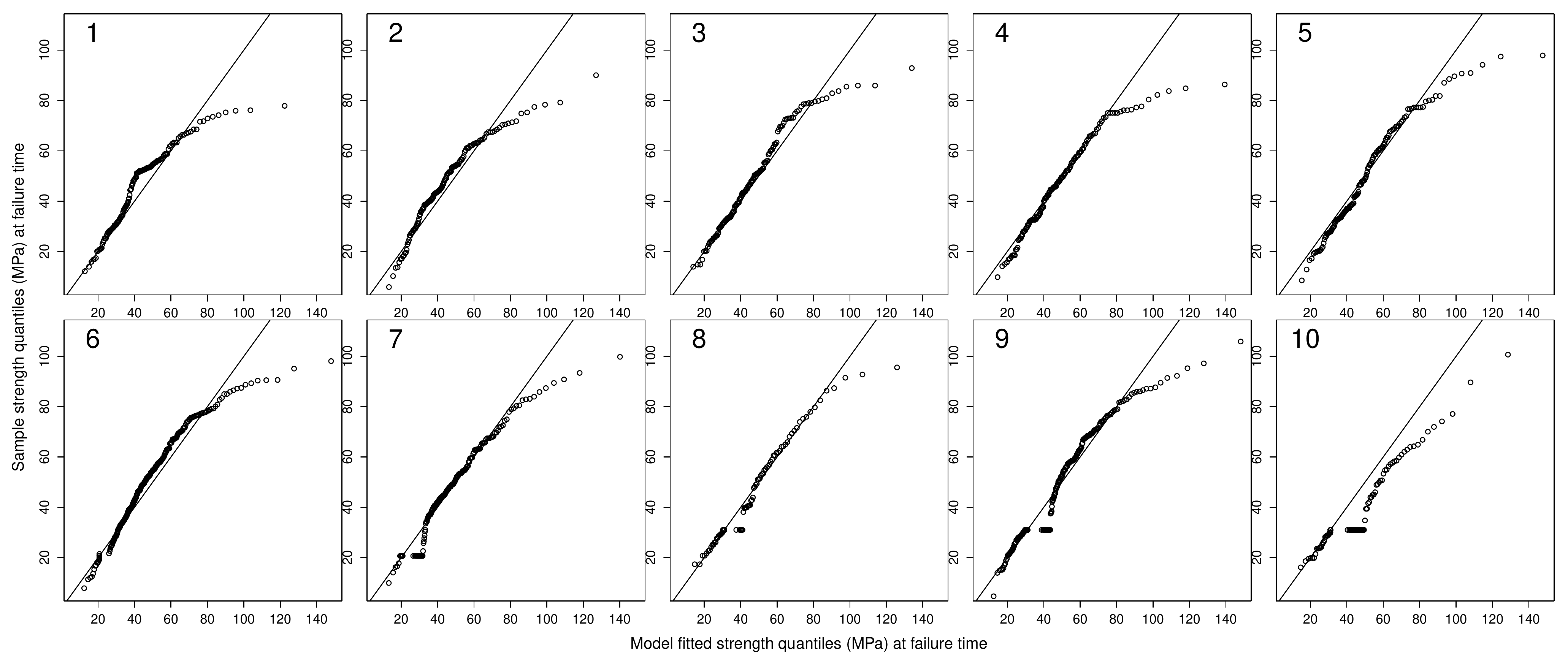}
	\caption{QQ plots for the fitted US model.  Each panel is labelled with the corresponding group number in the Forintek Canada dataset (see Tables \ref{tab:ramp} and \ref{tab:const}).}
	\label{fig:USQQ}
\end{figure}

%As noted in \cite{foschi1986duration2}, the use of expected Normal order statistics within the NLS fitting procedure is an approximation that may be incompatible with a few observations from the RCR data. % put in appendix? DONE
The median short-term strength of the sample (i.e., the median of group 3) is $\tau_M = 44.60$. 
Two points from the RCR data of group 6 had to be excluded from the fitting process due to incompatibility with the \cite{gerhards1987cumulative} NLS procedure (see Appendix \ref{app:US}).  By fitting the US model to all 10 groups, we obtain the parameter estimates $A = 68.46 (1.75)$, $B = 79.65 (1.91)$, and $w =0.4259  (0.0024)$, standard errors in brackets.
%that can cause a few data points from the RCR data to become out of bounds
% Later we want to assess whether these very small NLS SEs are reasonable or not.
%$B' = 0.01231 (0.00030)$
%> fit1s$coefficients[2] * tau0   
%[1] 79.64683
%> fit1s$coefficients[2,2] * tau0
%[1] 1.912523
%> tau0
%[1] 6467.75
Using these parameter estimates, we simulated 100,000 failure times from the US model for each of the 10 load profiles.  The QQ plots to visually assess the goodness-of-fit (as discussed in Section \ref{sec:GOF}) are shown in Figure \ref{fig:USQQ}, with each panel corresponding to the labelled group number.

The plots indicate that the fit is reasonable overall for the ROL data (groups 1-5); the deviations are primarily in the stronger pieces, where the model over-predicts their strength at failure.  The fit is less satisfactory  for the RCR data, as seen in the bottom panels (groups 6-10).  The horizontal sequence of points seen in panels 7 to 10 show that the US model tends to underestimate the number of failures that occur during the constant load period (where $\tau_c = 20.68$ for group 7, and $\tau_c=31.02$ for groups 8, 9, and 10).  The fit is poor for group 10, where the strengths of the constant-load survivors are largely over-estimated.  We examine the fits to the constant load period in more detail in section \ref{sec:modelcomp}.

The approximate value of the log-likelihood at the parameter estimates is 2960, by using KDE on the simulated failure times to obtain their probability density (see Section \ref{sec:GOF}). %This corresponds to a BIC of...

\subsection{Canadian model}
We ran the ABC-based MCMC model fitting procedure discussed in section \ref{est:canadian}, following these algorithmic settings as in \citet{yang2019bayesian}:  first, short tuning runs were used to determine an appropriate tolerance bandwidth of $\delta = 2.0$; then, we obtained our final 500 posterior samples of $\theta$ by using a burn-in length of 100,000 iterations and a thinning interval of 10,000.   For each of these 500 sampled values of $\theta$, we simulated 100,000 failure times from the Canadian model under the 10 different load profiles, and used these failure times to compute approximate log-likelihood values.  We take the sampled $\theta$ yielding the highest log-likelihood to be the parameter estimates, to facilitate our subsequent comparisons with other models. These are shown in Table \ref{tab:canfit} alongside 95\% credible intervals, obtained by taking the 0.025 and 0.975 quantiles of the 500 MCMC samples of $\theta$.

\begin{figure}[!htbp]
	\centering
	\includegraphics[scale=0.427]{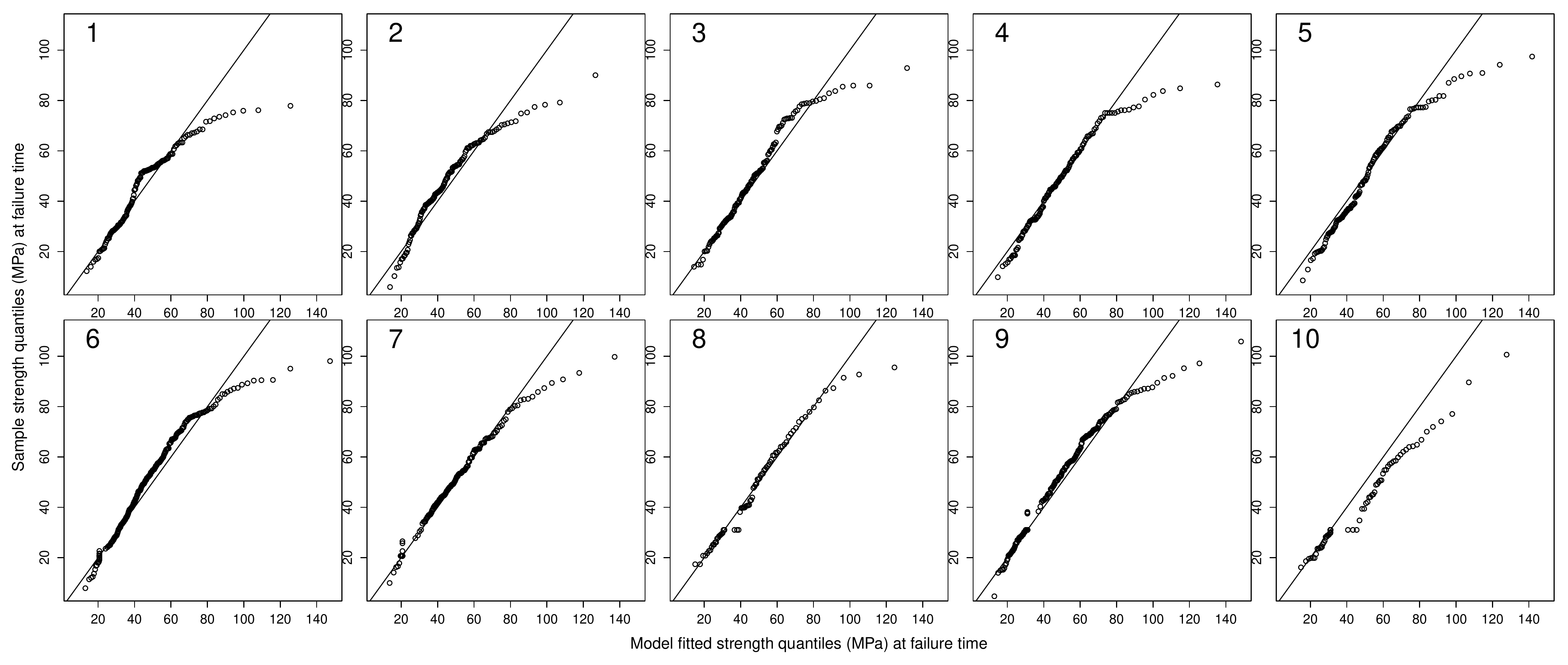}
	\caption{QQ plots for the fitted Canadian model.  Each panel is labelled with the corresponding group number in the Forintek Canada dataset (see Tables \ref{tab:ramp} and \ref{tab:const}).}
	\label{fig:CAQQ}
\end{figure}

%The distributions of the parameters are summarized in Table \ref{tab:canfit}:  
QQ plots to visually assess the goodness-of-fit (as discussed in Section \ref{sec:GOF}) are shown in Figure \ref{fig:CAQQ}, with each panel labelled with the corresponding group number.  The plots show fits to the ROL data (groups 1-5) that are nearly indistinguishable from the US counterpart in Figure \ref{fig:USQQ}.  In contrast, the fits to the RCR data (groups 6-10) are noticeably better than the US model, especially for groups 7 and 9; the deviation in the group 10 fit is also less pronounced.  The approximate value of the log-likelihood at the parameter estimates is 3131, by using KDE on the simulated failure times to obtain their probability density (see Section \ref{sec:GOF}).

\begin{table}[!htbp]
	\caption{Summary of fitted Canadian model parameters:  estimates and 95\% credible intervals based on the $0.025$ and $0.975$ quantiles of the MCMC samples. The notation $\mu_x$ and $\sigma_x$ denotes the lognormal mean and variance parameters associated with the random effect $x$ in the Canadian model (\ref{Can_model}). }
	\begin{center}
		\begin{tabular}{c|cc}
			Parameter &  Estimate   &           $95\%$ interval \\
			\hline
			$\mu_{a}$ &  -12.6  &    (-13.2, -12.2) \\
			$\sigma_{a}$ &   0.41  &     (0.16, 0.43) \\
			$\mu_{b}$ &   3.66  &     (2.99, 4.11) \\
			$\sigma_{b}$ &   0.09  &     (0.06, 0.30) \\
			$\mu_{c}$ & -46.4  & (-58.9, -13.0) \\
			$\sigma_{c}$ &   0.21 &     (0.06, 0.87) \\
			$\mu_{n}$ &  -1.89  &    (-2.38, 0.09) \\
			$\sigma_{n}$ &   0.33 &     (0.06, 0.55) \\
			$\mu_{\sigma_0}$ &  0.39  &    (-0.93, 0.90) \\
			$\sigma_{\sigma_0}$ &   0.15 &     (0.07, 0.50) \\
		\end{tabular}
	\end{center}
	\label{tab:canfit}
\end{table}

%-12.59 -13.15 -12.18
%-46.41 -58.87 -12.97
%             2.5% 97.5%
%V1   -7.61  -8.17 -7.21
%V2    0.41   0.16  0.43
%V3    3.66   2.99  4.11
%V4    0.09   0.06  0.30
%V5  -41.43 -53.90 -8.00
%V6    0.21   0.06  0.87
%V7   -1.89  -2.38  0.09
%V8    0.33   0.06  0.55
%V9    0.39  -0.93  0.90
%V10   0.15   0.07  0.50
%, as recommended in \citet{fearnhead2012constructing}.
%would give the desired ABC-MCMC acceptance rate of around 1\%
  % by adapting the software of \citet{yang2019bayesian}.

\subsection{Gamma process model}

We fit the gamma process model as discussed in section \ref{est:gamma}, following these algorithmic details as in \citet{wong2019duration}: load increments of $\tau_i - \tau_{i-1}=0.1379$  (20psi) for the sequence of load levels, uniform priors for MCMC sampling from the posterior distributions of the parameters, and 100,000 MCMC iterations.  To determine the number of time breakpoints for the piece-wise power law, the model was first fitted with no breakpoints, and then breakpoints were added one at a time until the BIC did not show further improvement.  For calculation of BIC, we used the parameters among the MCMC samples that had the highest log-likelihood.  We obtained BICs of -6085, -6172, -6185, and -6180, corresponding to 0, 1, 2, and 3 breakpoints respectively.  Thus, we included 2 breakpoints in our final model, corresponding to the lowest BIC value obtained and a log-likelihood of 3122.  %calculated exactly

As before, we take the sampled parameter values yielding the highest log-likelihood to be the estimate. These are shown in Table \ref{tab:gammafit} alongside 95\% credible intervals, obtained by taking the 0.025 and 0.975 quantiles of the MCMC samples.  Note that $a_1, a_2, a_3$ and $t_1, t_2$ are the powers and times associated with the 2 breakpoints; namely, the first piece of the power law is estimated as $g(t) = (t / 0.00144)^{3.7e-9}$ for $0 < t \le 0.00144$ hours.  We observe that $a_1$ is very close to zero, indicating that the model suggests no apparent DOL effect in the initial 0.00144 hours (approximately 5 seconds) of loading. Also of note is that there is no clear load threshold, as the credible interval for $\tau^*$ indicates the threshold can plausibly be zero.

\begin{figure}[!htbp]
	\centering
	\includegraphics[scale=0.427]{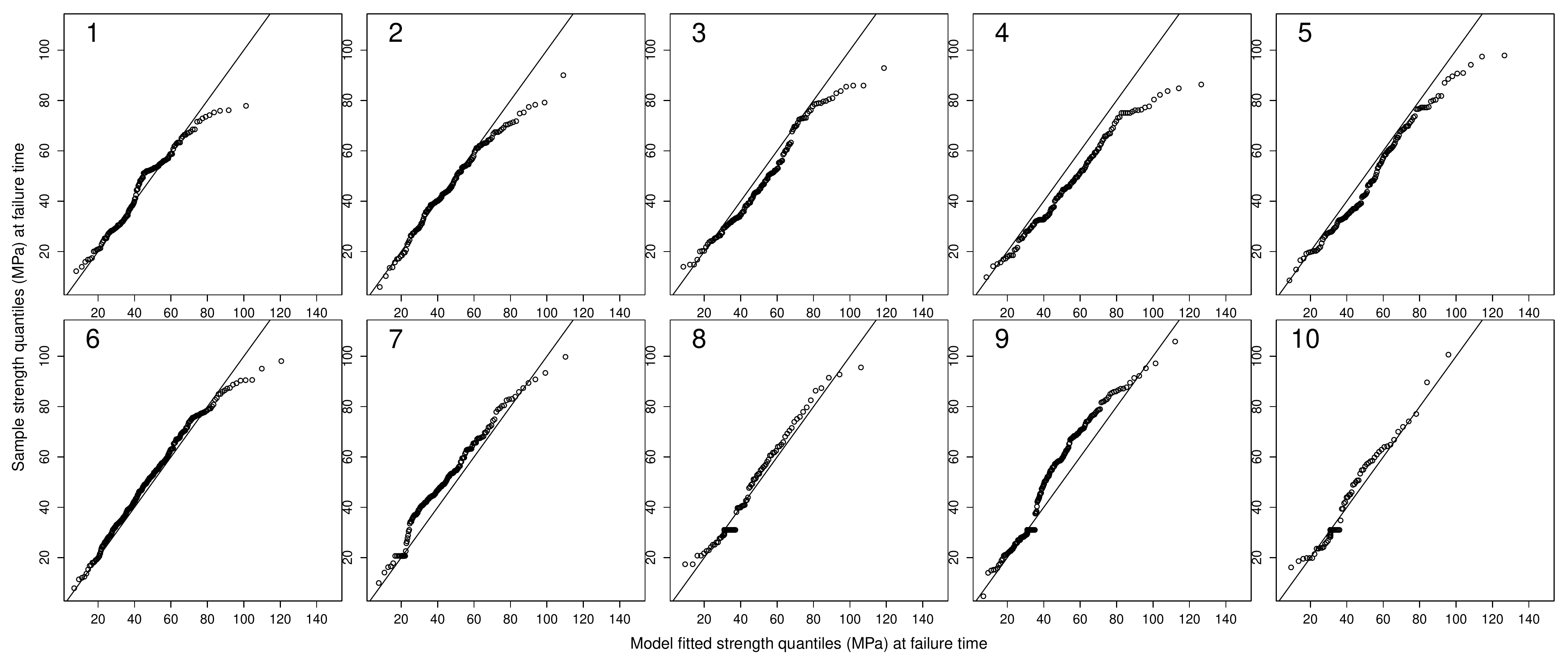}
	\caption{QQ plots for the fitted gamma process model.  Each panel is labelled with the corresponding group number in the Forintek Canada dataset (see Tables \ref{tab:ramp} and \ref{tab:const}).}
	\label{fig:GPQQ}
\end{figure}

QQ plots to visually assess the goodness-of-fit (as discussed in Section \ref{sec:GOF}) are shown in Figure \ref{fig:GPQQ}, with each panel labelled with the corresponding group number.  The plots have some notable differences compared to Figures \ref{fig:USQQ} and \ref{fig:CAQQ}.  First, the gamma process fits handle the strongest pieces well, resulting in much fewer outliers in all the groups.  Second, it relatively overestimates the strength of the group 4 ramp load.  Third, the model has mixed results with RCR data: it struggles with groups 7, 8, and 9, showing similar weaknesses as in the US model; however, it also provides the best visual fit to groups 6 and 10.

%; interestingly, that group has a lower mean strength than group 3, despite the faster loading rate.

% fits to the ROL data (groups 1-5) that are nearly indistinguishable from the US counterpart in Figure \ref{fig:GPQQ}.  In contrast, the fits to the RCR data (groups 6-10) are noticeably better than the US model, especially for groups 7 and 9; the deviation in the group 10 fit is also less pronounced.  

\begin{table}[!htbp]
	\caption{Summary of fitted Gamma process model parameters:  estimates and 95\% credible intervals based on the $0.025$ and $0.975$ quantiles of MCMC samples. }
	\begin{center}
		\begin{tabular}{c|cc}
			Parameter &  Estimate   &           $95\%$ interval \\
			\hline
			$u$ &  0.084  &    (0.077, 0.104) \\
			$a_1$ &   $3.7 \times 10^{-9}$  &     $(4.6 \times 10^{-14}, 2.1 \times 10^{-3})$ \\
			$a_2$ &   0.027  &     (0.018, 0.028) \\
			$a_3$ &   0.094  &     (0.054, 0.103) \\
			$t_1$ &   0.00144  &    (0.00015, 0.00493) \\
			$t_2$ &   2327 &     (289, 2890) \\
			$\tau^*$ &  4.35  &    (0, 4.45) \\
			$\xi$ &   0.27 &     (0.20, 0.30) \\
		\end{tabular}
	\end{center}
	\label{tab:gammafit}
\end{table}

%    g(sc, log_b, s1, log_s0, log_c, log_d, log_s2, log_s3) %=% theta
%# Note: s1 is u, s0 is tau* 
%> theta.summary
%[,1]         [,2]         [,3]         [,4]       [,5]       [,6]         [,7]      [,8]
%0.2737652 3.693296e-09 0.0005758969 6.304528e+02 0.02687230 0.09381795 0.0014384249 2326.8020
%2.5%  0.1983848 4.558155e-14 0.0005278884 1.233063e-17 0.01819593 0.05434965 0.0001498149  289.1655
%97.5% 0.2973699 2.093856e-03 0.0007141121 6.450817e+02 0.02833962 0.10259036 0.0049301468 2890.4086

% BIC values: -6085.165 -6171.722 -6184.962 -6180.161 (0 to 3 breakpoints)

\subsection{Comparing the models} \label{sec:modelcomp}

The three models considered differ in the number of estimated parameters:  the US model has three, the Canadian model has 10, while the gamma process model (with 2 breakpoints) has eight.  Naturally, more complex models  are expected to provide better fits to the data; thus, statistical criteria for model comparison such as BIC take into account the number of parameters estimated.  Using the log-likelihood values for each model, the calculated BICs for the US, Canadian, and gamma process models are -5898, -6188, and -6184, respectively (lower is better).  Equivalently, the likelihood ratio of the US model to either of the Canadian or gamma process models is $\approx 0$, after compensating for the differences in the number of parameters; the likelihood ratio of the Canadian model to the gamma process model is 55, after compensating for the two additional parameters in the Canadian model.  Thus according to this criterion \citep{kass1995bayes}, for these data the Canadian and gamma process models are very strongly preferred over the US model, while the Canadian model is somewhat preferred over the gamma process model.

\begin{figure}[!htbp]
	\centering
	\includegraphics[scale=0.427]{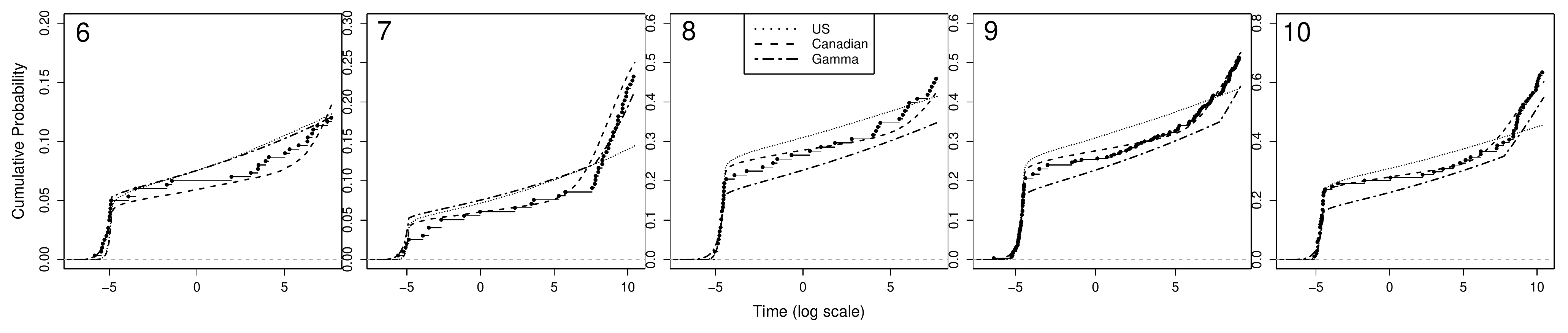}
	\caption{Failure time distributions of the constant load data (groups 6-10), plotted on the log scale.  The points indicate the observed failure times, while the lines compare the fits of the US, Canadian, and gamma process models.}
	\label{fig:constLoad}
\end{figure}

As a further visual comparison of the fitted models, we take a closer look at the constant load portion of the data in Figure \ref{fig:constLoad}.  Here, the empirical distributions of failures occurring before the end of the constant load period are plotted for groups 6-10, with failure time shown on the log scale.  The fitted distributions for the three models are superimposed as dotted and dashed lines.  It can be seen that the Canadian model best fits this aspect of the data, and most closely follows the empirical distributions. The US model does not have the flexibility to fit the non-linear shape of the empirical distributions in these plots, while the gamma process model can only partially model this aspect via its piece-wise power law.

%The US model does not have the flexibility to capture the uptick in failures as the, while the gamma process model can only partially capture that with a piece-wise power law.

% then use BIC to statistically compare 
% comment on fit to 5th percentiles, etc.

Finally, we note that none of the models uniformly provides the best fit to all 10 groups individually, as seen in the visual assessments in Figures 1 through 4.  The main appealing feature of the US model remains its simplicity.  Using the statistical techniques presented, it is now viable to apply the Canadian model to these datasets and achieve substantially better fits than its US counterpart within the ADM framework, especially for the constant load data.  The gamma process model differs conceptually from the ADMs by having a closed-form likelihood for failure times that bypasses the need for numerical simulation, and it also has power-law parameters that are easy to interpret.

% Each model has its limitations, as seen in the visual assessments in Figures 1 through 4. Nonetheless,

\section{Reliability analysis}

An important goal of damage modeling is to conduct reliability assessments of members under various stochastic loadings.  To illustrate this analysis using our three fitted models, we adopt procedures similar to those used in other recent studies for generating stochastic load profiles $\tau(t)$ \citep[e.g.,][]{li2016reliability,gilbert2019reliability,yang2019bayesian}.  The load profile is given by
\begin{equation}\label{eq:deadpluslive}
\tau(t)=\phi R_o\frac{\gamma\tilde{D}_d + \tilde{D}_l(t)}{\gamma\alpha_d  + \alpha_l}
\end{equation}
where $\phi$ is the performance factor applied multiplicatively to the characteristic strength $R_o$ of the lumber population.  We take $R_o = 20.68$, the approximate 5th percentile of the Hemlock species under consideration. The random variables $\tilde{D}_d$ and $\tilde{D}_l$ represent the standardized dead and live loads.  The dead-to-live load ratio $\gamma$ is taken to be 0.25, as used previously in \citet{li2016reliability} and \citet{yang2019bayesian}.  The values $\alpha_d = 1.25$ and $\alpha_l = 1.5$ are used in the National Building Code of Canada (NBCC) 2015 edition.  A reference time period of 50 years is used, as recommended in \citet{bartlett2003load}.

To complete the load specification, probability distributions are assigned to $\tilde{D}_d$ and $\tilde{D}_l(t)$.  The standardized dead load $\tilde{D}_d$ is assumed to follow a Normal distribution with mean 1.05 and SD 0.1 \citep{bartlett2003load}, and fixed for the lifetime of the structure.  The live load is assumed to vary over time, consisting of a sustained component and an extraordinary component, so that $\tilde{D}_l(t) = \tilde{D}_s(t) + \tilde{D}_e(t)$ \citep{foschi1989reliability,gilbert2019reliability}.  The specific parameters for these components are adopted from the residential load profile in \citet{foschi1989reliability}. $\tilde{D}_s(t)$ consists of exponentially distributed periods with mean duration 10 years, and the size of the load in each period is generated from a gamma distribution with shape parameter 3.122 and scale parameter 0.0481.  Meanwhile, $ \tilde{D}_e(t)$ consists of brief periods with extraordinary loads, where each has exponential duration with mean 2 weeks and gamma distributed size  with shape parameter 0.826 and scale parameter 0.1023; the time between occurrences of extraordinary loads is also exponentially distributed with mean one year.

\begin{figure}[!htbp]
	\centering
	\includegraphics[scale=0.5]{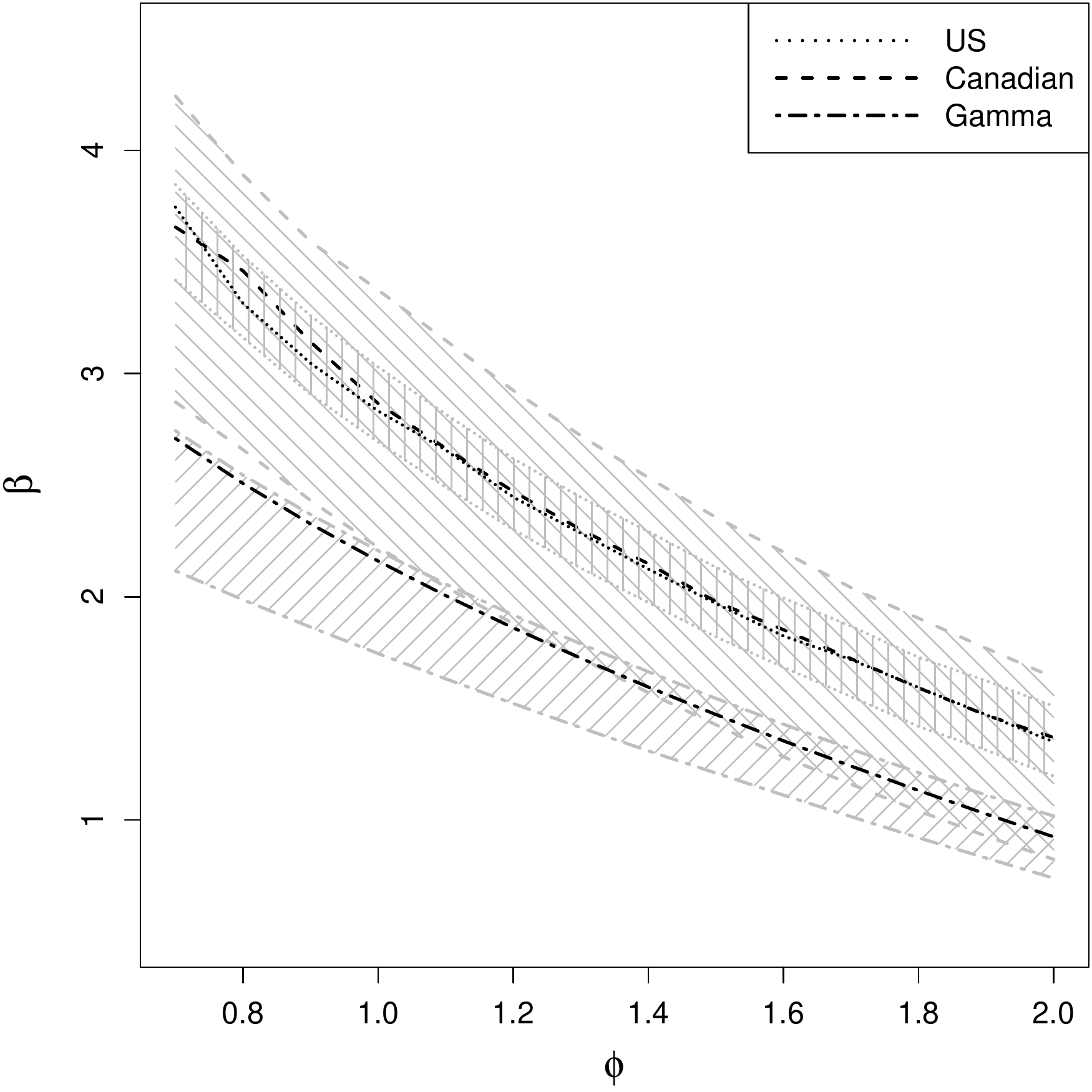}
	\caption{Reliability analysis results under residential loads.  The black lines show the $\phi-\beta$ relationship for the fitted US, Canadian, and gamma process models; the corresponding grey bands are 95\% intervals resulting from the uncertainties associated with the parameter estimates.}
	\label{fig:reliability}
\end{figure}

Under the first order reliability method \citep[FORM, see ][]{madsen2006methods}, the reliability index $\beta$ for a given $\phi$ is then calculated as $\beta = -\Phi^{-1}(p_f)$, where  $\Phi$ is the standard Normal cumulative distribution function and $p_f$ is the probability of failure based on a large number $N_R$ of simulated load profiles in equation (\ref{eq:deadpluslive}).  Using our parameter estimates for the three fitted models, we compute $\beta$ over a range of $\phi$ values, using $N_R = 100,000$ for each value of $\phi$.  The resulting curves are shown in Figure \ref{fig:reliability}, along with 95\% interval bands that account for uncertainty in the parameter estimates.  For the Canadian and gamma process models, the 95\% credible limits are obtained by calculating $\beta$ with the MCMC samples of the parameters.  To obtain the 95\% confidence limits for the US model, we assume the parameters are approximately normally distributed with mean and SD given by their estimates and standard errors from the NLS procedure.
%For a range of values of $\phi$,
%0.826,0.1023
% NOTES
% Cite svenson 2011?  different approach to likelihood estimation on constant load.  ABOVE in CANADIAN MODEL description?
% Dead load  N(1.05, 1) from NBCC 2015 & Bartlett 2003
% 1.25, 1.5 from NBCC 2015
% dead-to-live load ratio 0.25 as used in previous pubs Li & Lam
% Live load params from Foschi & Yao. and our Technometrics paper.
% 50 years: Bartlett
% First order reliability method Madsen 2006.  phi-beta relationship

% US model: assume pars are approx normally distributed with SE

We find that the estimated reliability indices based on the fitted US and Canadian models are remarkably similar for this example, given the differences in the quality of their fits to the Forintek dataset.  The Canadian model suggests that there is substantially more variability due to parameter uncertainty, compared to the US model estimated via NLS.  The gamma process model predicts a higher probability of failure than both ADMs.  However, after accounting for parameter uncertainty, the 95\% bands for the Canadian and gamma process models overlap for all $\phi\ge 1.0$.
%; though, we also emphasize the US and Canadian estimation techniques also differ.

\section{Discussion and conclusions}

In this paper, we presented three damage models -- the US ADM, Canadian ADM, and a gamma process model -- and associated statistical methods to fit these models to experimental data.  We considered the ramp load, constant load, and ramp-constant-ramp load test scenarios that have been previously used to assess the effects of load duration and rate on wood products. We showed how to extend existing methodology to fit failure time data from these scenarios in a coherent statistical framework that quantifies estimation uncertainty.   The models and methods were demonstrated via a novel analysis of a Hemlock dataset and an application to reliability analysis was made.

% we've illustrated the kind of analyses that can now be done with statistical machinery
This work illustrates the powerful analyses that can now be carried with modern statistical computation.  We anticipate that the methods presented can be useful to practitioners working with DOL and ROL data, for example, from experiments involving new engineered wood products.  To facilitate the use of these methods, the programs and code for carrying out these analyses are provided in the Github repository \url{https://github.com/wongswk/damage-models}.

We note that there is also practical utility from analyzing DOL and ROL data with multiple models.  Since all models are approximations, there is model uncertainty associated with conducting reliability analyses over a 50-year period based on accelerated testing.  This is in addition to uncertainties in the parameter estimates that are quantified by the methods presented.  Thus, it can be helpful to examine the range of possible reliability outcomes across different models, and potentially combine the results using some expert judgment.   
Finally, while our analyses in this paper were entirely data-driven, one may also incorporate expert knowledge at the model fitting stage, by imposing priors or constraints on model parameters. 

%associated with fitting the models -- which the methods 
% extrapolating estimates based on accelerated testing to reliability analyses over a 50-year period can be sensitive to model assumptions.  Along with principled assessment of statistical uncertainty, 
% suite of models gives broad range of results, some non-overlapping.  could use some expert judgment, or in some way weight the different models.  + uncertainty assessment.  each has strengths in goodness of fit.
% completely data driven, can add expert knowledge  
%principled assessment of uncertainty

\section*{Acknowledgements}
The work reported in this paper was partially supported by FPInnovations and a CRD grant from the Natural Sciences and Engineering Research Council of Canada.  
The author is greatly indebted to Conroy Lum and Erol Karacabeyli from FPInnovations for introducing the author to this important area of research, sharing extensive advice during the conduct of this study, and providing the Forintek dataset analyzed herein.
The author also thanks James V Zidek for helpful discussions during the preparation of the manuscript.
%%The author thanks Erol Karacabeyli from FPInnovations for providing the Forintek dataset analyzed in this paper

\appendix

\section{US model fitting details} \label{app:US}
The analytic solutions of the time-to-failure for the US model are as follows, under the three test profiles considered (where $B' = B / \tau_M$):
\begin{enumerate}
	\item[R.] $T_f = [\exp(wZ) / (B'k)] \log \{[ B'k/\exp(wZ)] \exp(A) +1 \}$
	\item[C.] $T_f = \tau_c / k - \exp(wZ) / (B'k) + \{ \exp[ - B' \tau_c / \exp(wZ) ] \} [ \exp(wZ)/(B'k) + \exp(A)]$ %$T_f = \exp [A - B' \tau_c / \exp(wZ)]$
	\item[RCR.] $T_f - T_1 = [\exp(wZ) / (B'k)] \log \{[ B'k/\exp(wZ)] \exp(A) ( 1 - \alpha(T_1)) +1 \}$, where $\alpha(T_1) = (T_1-T_0) \exp [ -A + B' \tau_c / \exp(wZ)] + \exp(-A) \exp(wZ) / (B'k) \{ \exp [B'k/\exp(wZ)] - 1  \}  $ is the damage sustained by the end of the constant load test.
\end{enumerate}
The expressions for R and C are also given in \cite{gerhards1987cumulative}, equations 5 and 6.

%For each specimen, the dependent variable for nonlinear least squares is taken to be the logarithm of the LHS of the equation (T1, T2 or T3) corresponding to the scenario under which the failure occurred.

For each specimen, the difference used for nonlinear least squares is log(LHS)$-$log(RHS) from the equation (R, C, or RCR) corresponding to the scenario when failure occurred; the value substituted for $Z$ is the expected value of its standard Normal order statistic within that sample.   As noted in \cite{gerhards1987cumulative}, this substitution is an approximation that may be incompatible with a few observations by causing the RHS of the RCR equation to involve the logarithm of a negative value; such data points are excluded from the fit. % put in appendix?
To account for the time scale variation between ramp and constant load and maintain homoscedasticity, the residuals for the constant load data are reweighted by a factor of $1/(B' \tau_c)$ using the current estimate of $B'$.  Parameter estimation and reweighting is done iteratively until convergence.  See \cite{gerhards1987cumulative} for additional details.

% $\log(T_f)$ when failure occurs during ramp or constant load, and $\log(T_f-T_1)$ for scenario T3 to ensure comparable scaling.  The 

\section{Canadian model fitting details} \label{app:Can}

As in \cite{yang2019bayesian}, the random effect distributions are assumed to be as follows:
\begin{eqnarray*} \label{eqn:random-effect}
a|\mu_{a},\sigma_{a} & \sim & \text{Log-Normal}(\mu_{a},\sigma_{a}) \nonumber \\
b|\mu_{b},\sigma_{b} & \sim & \text{Log-Normal}(\mu_{b},\sigma_{b}) \nonumber \\
c|\mu_{c},\sigma_{c} & \sim & \text{Log-Normal}(\mu_{c},\sigma_{c})\\
n|\mu_{n},\sigma_{n} & \sim & \text{Log-Normal}(\mu_{n},\sigma_{n}) \nonumber \\
\eta|\mu_{\sigma_{0}},\sigma_{\sigma_{0}} & \sim & \text{Log-Normal}(\mu_{\sigma_{0}},\sigma_{\sigma_{0}}) \text{ and set } \sigma_{0} = \frac{\eta}{1+\eta}.\nonumber
\end{eqnarray*}

As shown in \cite{wong2018dimensional} and \cite{yang2019bayesian} and  reproduced here to aid exposition, the solution $T_s$ for time-to-failure under a ramp-load test with  $k = k_s$ is determined by the equation
\begin{equation}\label{eq:Tssoln}
H(T_{s})=\frac{\left(akT_{s}\right)^{b}}{\left(ckT_{s}\right)^{n(b+1)/(n+1)}}\left(\frac{\mu(n+1)}{T_{s}}\right)^{\frac{b-n}{n+1}}\int_{0}^{-\log H(T_{s})}e^{-u}u^{(b+1)/(n+1)-1}du
\end{equation}
where $H(t)$ is the integrating factor in equation (\ref{eq:intfac}).

%\begin{equation} \label{eq:intfac}
%H(t)=\exp\left\{ -\frac{1}{\mu}\left(ckT_{s}\right)^{n}\frac{T_{s}}{n+1}\left(\frac{t}{T_{s}}-\sigma_{0}\right)^{n+1}\right\} .
%\end{equation}
Thus $T_s$ (and hence $\tau_s = k_s T_s$) is implicitly a function of $a$, $b$, $c$, $n$, $\sigma_0$; given the values of these random effects, we first compute $\tau_s$.   Further, a constant--load test that sets $k=k_s$ for the initial ramp-loading portion ($t \le T_0$) then has a failure time $T_c$ that can be expressed in terms of $T_s$ and the random effects, namely
\[
T_{c}=-\frac{1}{C_{2}}\log\left(\frac{\frac{C_{1}}{C_{2}}H^{\star}(T_{0})+C_{3}}{1+\frac{C_{1}}{C_{2}}}\right)
\]
where
\begin{eqnarray*}
	C_{1} & = & \frac{1}{\mu}\left[akT_{s}\left(\frac{T_{0}}{T_{s}}-\sigma_{0}\right)\right]^{b}\\
	C_{2} & = & \frac{1}{\mu}\left[ckT_{s}\left(\frac{T_{0}}{T_{s}}-\sigma_{0}\right)\right]^{n}\\
	C_{3} & = & \alpha(T_{0})H^{\star}(T_{0})\\
	H^{\star}(T_{0}) & = & \exp\left\{ -C_{2}T_{0}\right\} \\
	\alpha(T_{0}) & = & \frac{1}{H(T_{0})}\frac{\left(akT_{s}\right)^{b}}{\left(ckT_{s}\right)^{n(b+1)/(n+1)}}\left(\frac{\mu(n+1)}{T_{s}}\right)^{\frac{b-n}{n+1}}\int_{0}^{-\log{H(T_{0}})}e^{-u}u^{(b+1)/(n+1)-1}du.
\end{eqnarray*}

\textbf{RCR solution.} Continuing from this result, we need the damage sustained $\alpha(T_1)$ for pieces that have not failed by the end of the constant load test.  During the constant load period $T_0 \le t \le T_1$ when $\tau(t) = \tau_c$, we have
%\begin{eqnarray*}
$\frac{d}{dt} \left[e ^ {-C_2 t} \alpha(t) \right] = C_1 e^{-C_2 t}$
%\end{eqnarray*}
and so integration yields
\begin{eqnarray*}
%e^{-C_2 T_1} \alpha(T_1) =  e^{-C_2 T_0} \alpha(T_0)  + \frac{C_1}{C_2} \left(  e^{-C_2 T_1} -  e^{-C_2 T_0} \right)
\alpha(T_1) =  e^{C_2 (T_1-T_0) } \alpha(T_0)  + \frac{C_1}{C_2} \left( 1 -  e^{C_2 (T_1-T_0) } \right).
\end{eqnarray*}
When $\alpha(T_1) < 1$, the RCR test applies $\tau(t) = k(t-T_1)$ for $t > T_1$ with $k=k_s$ to yield failure time $T_f$.  Using the same integrating factor defined in equation (\ref{eq:intfac}), we now have for  $t > T_1$
  \begin{eqnarray*}
	\frac{d}{dt} \left[ \alpha(t) H(t-T_1) \right]  = \frac{1}{\mu} \cdot H(t-T_1) \left[ {a} k T_s  \left( \frac{t-T_1}{T_s} - \sigma_0 \right) \right]^b.
\end{eqnarray*}
Damage begins accumulating again at $t = T_1 + \sigma_0 T_s$, so integrating over $T_1 \le t \le T_f$ we obtain
\begin{eqnarray*}
	\alpha(T_f)H(T_f-T_1) - \alpha( T_1 )  = \int_{\sigma_0 T_s}^{T_f - T_1} \frac{1}{\mu} \cdot H(t) \left[ {a} k T_s  \left( \frac{t}{T_s} - \sigma_0 \right) \right]^b \,dt.
\end{eqnarray*}
Finally the change of variables $u = -\log H(T_f - T_1)$ yields  equation (\ref{eq:TfRCRsoln}) that can be solved for $T_f$.
%an equation analogous to (\ref{eq:Tssoln}) 

\textbf{ROL solution.}  Now we consider ramp load tests with varying rates of loading $k$.  After solving for $\tau_s$, write the integrating factor
\begin{equation*} \label{eq:intfacROL}
\tilde{H}(t)=\exp\left\{ -\frac{1}{\mu}\left(c \tau_s \right)^{n}\frac{\tau_s  }{k(n+1) }\left(\frac{kt}{ \tau_s }-\sigma_{0}\right)^{n+1}\right\} .
\end{equation*}
Then under $\tau(t) = kt$, the failure time $T_f$ is the solution to this equation, which may be derived analogously to (\ref{eq:Tssoln}),

\begin{equation*}\label{eq:TsolnROL}
\tilde{H}(T_{f})=\frac{\left(a \tau_s \right)^{b}}{\left(c \tau_s \right)^{n(b+1)/(n+1)}}\left(\frac{\mu k (n+1)}{ \tau_s }\right)^{\frac{b-n}{n+1}}\int_{0}^{-\log \tilde{H}(T_{f})}e^{-u}u^{(b+1)/(n+1)-1}du.
\end{equation*}

%TeqnR <- function( Tf, theta ) {
%	with(as.list(theta),{
%		Ts <- tau_s/k_s
%		A <- a*tau_s
%		C <- c*tau_s
%		intfac <- 1/mu * (c*tau_s)^n * (tau_s/k) /(n+1) * (Tf/(tau_s/k)-s0)^(n+1)
%		A^b / C^(n*(b+1)/(n+1)) * (mu*k/tau_s * (n+1))^((b-n)/(n+1))* pgamma(intfac, (b+1)/(n+1)) * gamma((b+1)/(n+1)) - exp(-intfac)
%	})
%}

% and integrating (\ref{Can_model}) over  yields

%Thus the complete solution for the constant-load failure time $T$ is
%\begin{equation} \label{eqn:sol}
%T=\begin{cases}
%T_{s} & \text{if }T_{s} \leq T_{0}\\
%T_{c} & \text{if }T_{s}>T_{0}
%\end{cases}.
%\end{equation}

\bibliographystyle{apalike}
\bibliography{rol-survivor}

\end{document}